\documentstyle[aps,prd,floats]{revtex}

\newcommand{\beq}{\begin{equation}}
\newcommand{\eeq}{\end{equation}}
\newcommand{\bea}{\begin{eqnarray}}
\newcommand{\beas}{\begin{eqnarray*}}
\newcommand{\eea}{\end{eqnarray}}
\newcommand{\eeas}{\end{eqnarray*}}
\newcommand{\ba}{\begin{array}}
\newcommand{\ea}{\end{array}}

\begin{document}
\draft
\input epsf 

\title{Inflatonic Q-ball evaporation: A new paradigm for reheating the 
Universe}

\author{Anupam Mazumdar}

\address{The Abdus Salam, International Centre for Theoretical Physics,
Starda Costiera-11, 34100, Trieste, Italy\\
and \\
Physics Department, McGill University,3600-University Road, Montreal, H3A 2T8,
Canada}

\maketitle
\begin{abstract}
We show that the inflaton condensate associated with a global symmetry 
can fragment into quasi stable Q balls after the end of inflation, 
provided the inflaton oscillations give rise to an effective equation 
of state with negative pressure. We argue that chaotic inflation with 
a running inflaton mass may give rise to the desired scenario where the 
process of fragmentation into inflatonic Q balls can actually take place 
even though there is total zero charge. We show that such inflatonic Q balls
will reheat the Universe to a sufficiently low temperature via surface
evaporation.
\end{abstract}

\vskip2pc
\section{Introduction}
\vspace{-.3truecm}

Inflation, as the early Universe paradigm, inflates away all spatial 
inhomogeneities except the quantum fluctuations which is observationally 
bounded by one part in $10^{5}$, see \cite{liddle-lyth00}. The creation 
of matter and entropy generation takes place only when the inflaton, 
as a condensate, breaks-up and decays into the Standard Model degrees 
of freedom. This may either happen via perturbative decay of the condensate 
\cite{albrecht82,kolbturner90} or via non-perturbative processes dubbed 
as preheating~\cite{traschen90,kofman94}. Preheating typically involves 
an amplification in some of the fluctuation modes as well as the 
fragmentation of the inflaton condensate. Reheating dynamics depends 
very much on the form assumed for the inflaton potential, and for some 
choices, the inflaton condensate may also form $Q$-balls, see 
\cite{enqvist02a,enqvist02b,maz}. The $Q$-balls do not decay throughout their
volume, but evaporates only via surface interaction \cite{cohen86},
see a review \cite{mazumdar02} on $Q$-balls and their cosmological 
consequences.

\vspace{-.3truecm}

\section{Reheating as a surface effect}
\vspace{-.3truecm}

Usually the process of reheating is taken to be entirely a volume
effect. This can be problematic, however, especially if the scale
of inflation is high, i.e. $H_{inf}\sim 10^{15}-10^{16}$~GeV, as it
is sometimes assumed in order to provide the right magnitude for the
density perturbations, and also for reasons that have to do with
non-thermal heavy dark matter production or exciting right-handed
Majorana neutrinos for leptogenesis, etc., (see for alternative
low scale non-thermal leptogenesis \cite{allahverdi02})

As it is well known, the entropy thus dumped into the Universe
may pose a problem for big bang nucleosynthesis by overproducing
gravitinos from a thermal bath; an often quoted bound on the reheat
temperature is $T_{\rm rh} \leq 10^9~{\rm GeV}$ \cite{ellis85b}.
Note that gravitino can also be produced non-thermally
\cite{maroto99}, and there are other supersymmetric relics such as 
inflatino \cite{nilles01}. Though, it was recently pointed out that 
inflatino production is not dangerous for the big bang nucleosynthesis 
\cite{allahverdi01}. Nevertheless, Obtaining such a low reheat 
temperature is a challenge for high scale inflation models especially
if the inflaton sector and the Standard Model sector are strongly coupled. 
One way to solve the gravitino problem is to dilute them via a brief 
period of late thermal inflation \cite{lythstewart}, or assuming that the
inflaton sector is a hidden sector while matter along with density 
perturbations are created from the observable sector Minimal supersymmetric 
Standard Model flat directions, see \cite{enqvist02c}. Low scale inflation 
can also solve the gravitino problem \cite{mazumdar99}.

A novel way to avoid the gravitino problem is reheating via the surface 
evaporation of an inflatonic soliton. Compared with the volume driven 
inflaton decay, the surface evaporation naturally suppresses the decay 
rate by a factor
\begin{equation}
\label{sur}
\frac{{\it area}}{{\it volume}} \propto L^{-1}\,,
\end{equation}
where $L$ is the effective size of an object whose surface is evaporating.
The larger the size, the smaller is the evaporation rate, and therefore
the smaller is the reheat temperature.

Reheating as a surface phenomenon has been considered
\cite{enqvist02a,enqvist02b} in a class of chaotic inflation models
where the inflaton field is not real but complex. As the inflaton
should have coupling to other fields, the inflaton mass should in
general receive radiative corrections \cite{liddle-lyth00}, resulting
in a running inflaton mass and in the simplest case in the inflaton
potential that can be written as
\begin{equation}
    \label{qpotr}
    V = m^2 |\Phi|^2
    \left[ 1 + K\log\left(\frac{|\Phi|^2}{M^2}\right)\right ]\,,
\end{equation}
where the coefficient $K$ could be negative or positive, and $m$ is
the bare mass of the inflaton. The logarithmic correction to
the mass of the inflaton is something one would expect
because of the possible Yukawa and/or gauge couplings to other fields.
Though it is not pertinent, we note that the potential Eq.~(\ref{qpotr})
can be generated in a supersymmetric theory if the inflaton has a
gauge coupling \cite{enqvist99,kasuya0061,enqvist0163} where
$K\sim -(\alpha/8\pi)(m_{1/2}^2/m_{\widetilde{\ell}}^2)$,
where $m_{1/2}$ is the gaugino mass and $m_{\widetilde{\ell}}$ denotes the
slepton mass and $\alpha$ is a gauge coupling constant. It is also
possible to obtain the potential Eq.~(\ref{qpotr}) in a non-supersymmetric 
(or in a broken supersymmetry) theory, provided the fermions live in a 
larger representation than the bosons. In this latter situation the value 
of $K$ is determined by the Yukawa coupling $h$ with 
$ K =-C({h^2}/{16\pi^2})$, where $C$ is some number.

As long as $|K| \ll 1$, during inflation the dominant contribution to the
potential comes from $m^2|\Phi|^2$ term, and inflationary slow roll
conditions are satisfied as in the case of the standard chaotic model.
COBE normalization then implies $m\sim 10^{13}$~GeV
\cite{enqvist02a,enqvist02b}. If $K < 0$, the inflaton condensate feels 
a negative pressure (see, \cite{mazumdar02} for the discussion on negative 
pressure of the MSSM condensate) and it is bound to fragment into lumps 
of inflatonic matter. Moreover, since the inflation potential 
Eq.~(\ref{qpotr}) respects a global $U(1)$ symmetry and since for a 
negative $K$ it is shallower than $m^2|\Phi|^2$, it admits a $Q$-ball 
solution \cite{mazumdar02}. Comparing with $Q$-balls along the MSSM 
flat directions, here the major difference is that the inflatonic condensate
has no classical motion along the imaginary direction as usually required
for a $Q$-ball solution.

\vspace{-.3truecm}
\section{$Q$-balls from the inflaton condensate}
\vspace{-.3truecm}

\begin{figure}[t!]
\centering
\hspace*{-7mm}
\leavevmode\epsfysize=5cm \epsfbox{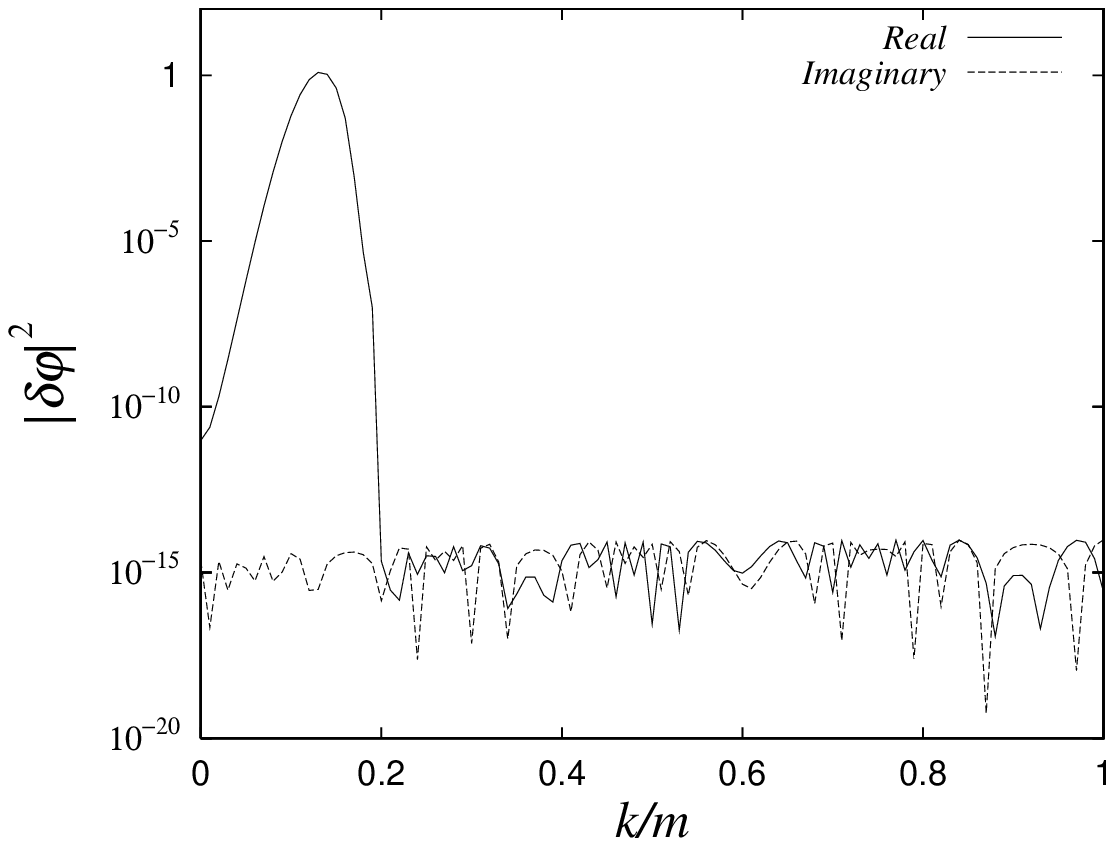}
\leavevmode\epsfysize=5cm \epsfbox{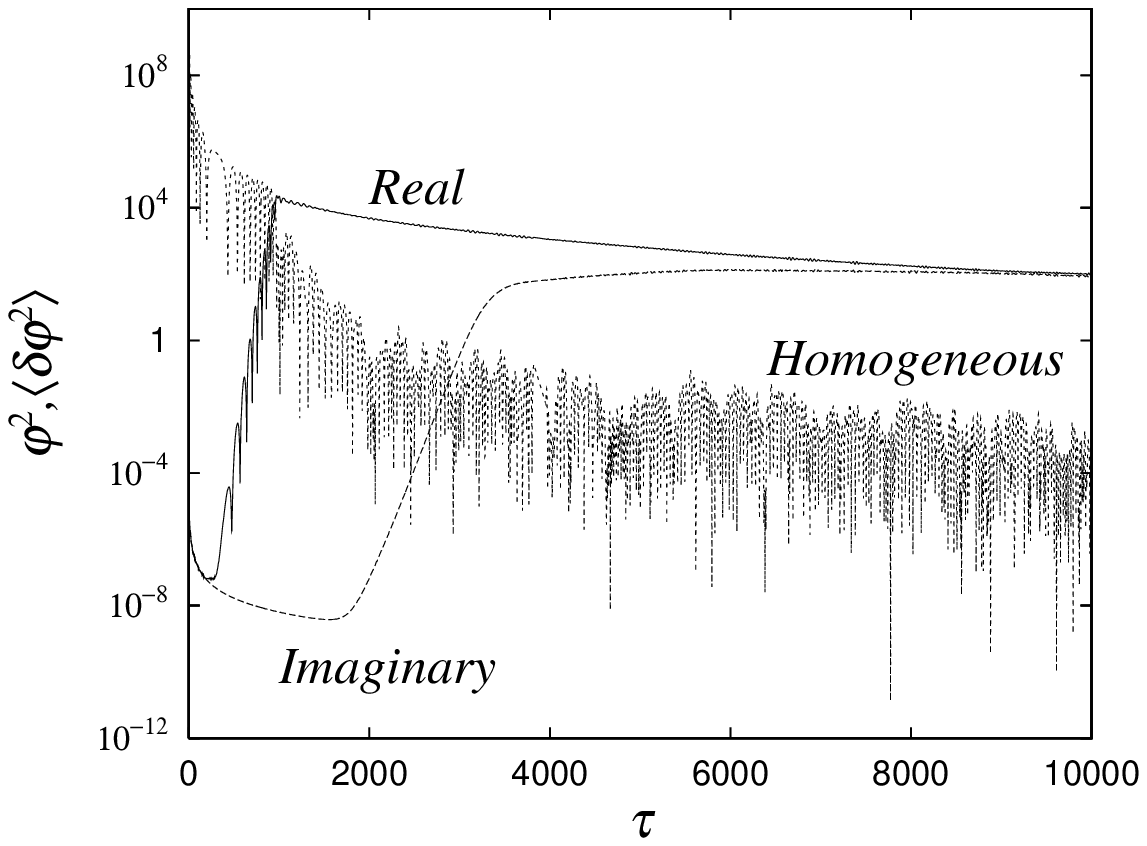}\ \nonumber \\
\leavevmode\epsfysize=5cm \epsfbox{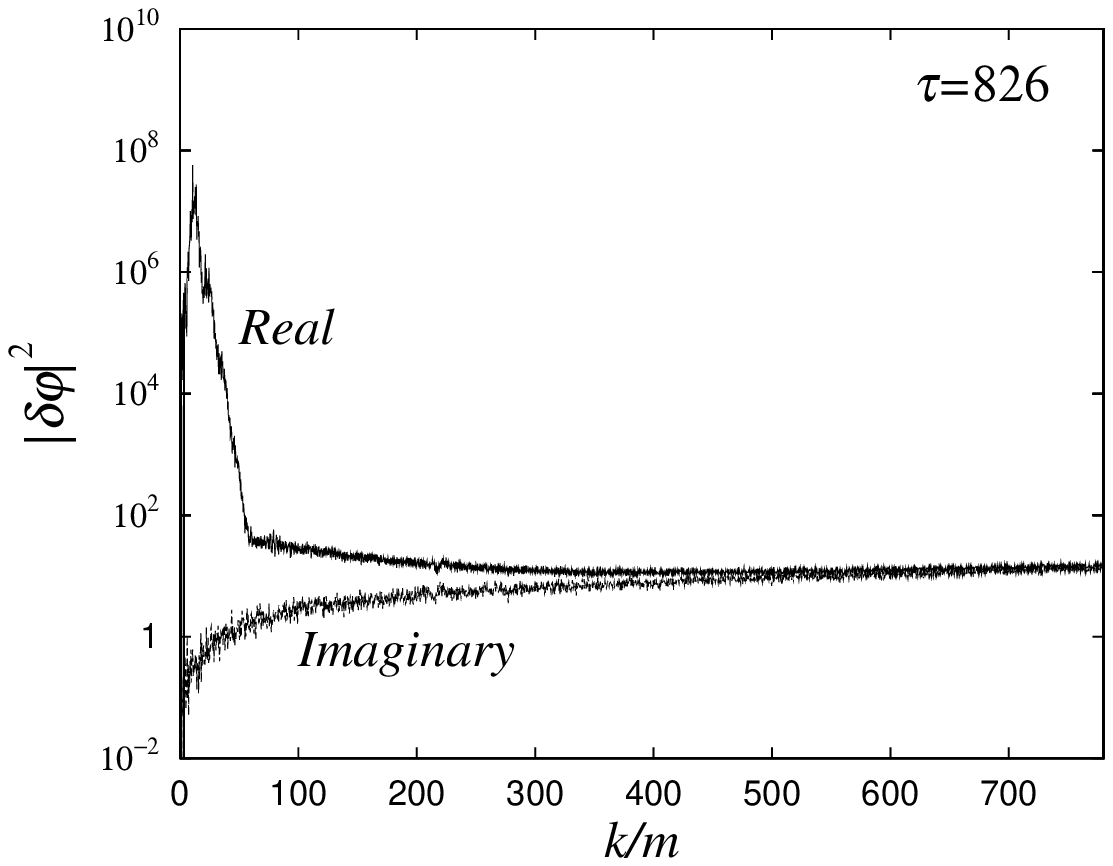}\\[2mm]
\caption{\label{mode-osc}
\small The first plot from left shows the instability bands
of the homogeneous mode of the inflaton along the real (solid) and imaginary
(dotted) directions. The second plot shows the result of
lattice simulation in the real and imaginary
directions, together with the evolution of the homogeneous  mode.
The third plot shows the power spectra of fluctuations at late times
All plots assume $K=-0.02$.}
\end{figure}

\begin{figure}[t!]
\centering
\hspace*{-7mm}
\leavevmode\epsfysize=6cm \epsfbox{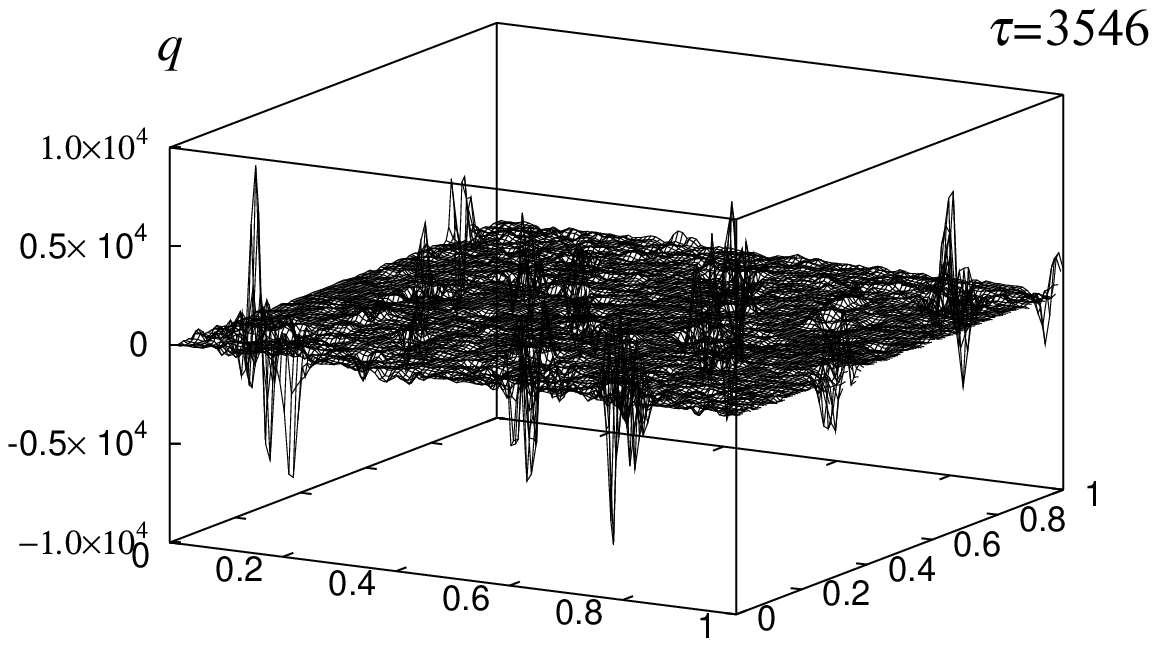}
\leavevmode\epsfysize=6cm \epsfbox{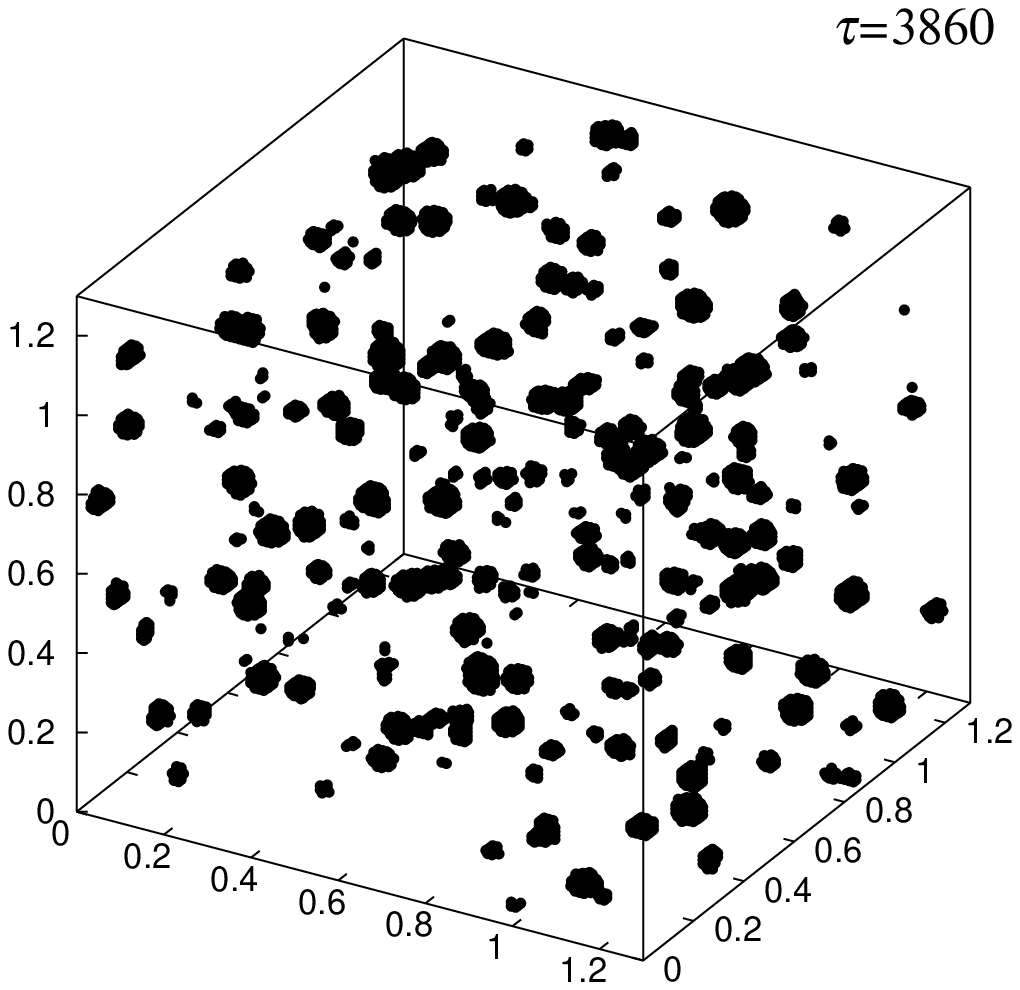}\\[2mm]
\caption{\label{3Dinf}
\small The first plot shows the charge density distribution in a small
sub-lattice at late times. The second plot shows inflatonic solitons
forming in $3D$ lattice.  Here $K=-0.02$}
\end{figure}

There are quantum fluctuations along both the real and imaginary 
directions which may act as the initial seed that triggers on the 
condensate motion in a whole complex plane \cite{enqvist02a,enqvist02b}.
The fluctuations in the real direction grow and drag the imaginary direction
along via mode-mode interactions, as illustrated by $2$ dimensional
lattice simulation in \cite{enqvist02b}, see Figs.~(\ref{mode-osc}).
The first plot shows the linear fluctuations without rescattering effects;
scattering effects are accounted for in the second plot. The late time
formation of inflatonic solitons is shown in Fig.~(\ref{3Dinf}). $Q$-balls
were observed to form with both positive and negative charges, as can be
seen in the first plot of Fig.~(\ref{3Dinf}), while keeping the net
global charge conserved. Inflatonic $Q$-balls are of same size because
the running mass potential resembles the MSSM flat direction potential
in the gravity mediated case, where the $Q$-ball radius is independent
of the charge.


$Q$-balls of size $R \sim |K|^{-1/2}m^{-1}$ form when the
fluctuations grow nonlinear \cite{enqvist02a,enqvist02b}.
Since the growth rate of fluctuation is $\sim |K|m$, the Hubble 
parameter at the formation time can be estimated as 
$H_{f} \sim \gamma |K|m$, where $\gamma$ is a numerical coefficient 
less than one. For $|K|\ll 1$, we can approximate the decrease in 
the amplitude of the oscillations by $\phi_f \sim \phi_i(H_f/H_i)$ as 
in the matter dominated era, where $\phi_i \simeq M_{\rm P}/\sqrt{12\pi}$ 
denotes the amplitude at the end of inflation in the chaotic model, and 
$H_i\sim m$ when the oscillations begin. The total charge of a $Q$-ball 
is given 
by $Q\sim(4\pi/3)R^3n_q\sim(1/9)\beta\zeta^2\gamma^2|K|^2R^3mM_{\rm P}^2$,
where $n_q = \beta \omega\phi_0^2$, $\phi_0 \simeq \zeta \phi_f$, and
$\beta \ll 1$ and $\zeta \geq 1$ are numerical factors. Note that the
$Q$-balls have very small relative velocity and their interaction does not
necessarily give rise to a violent annihilation. The $Q$-balls and 
anti-$Q$-balls rather merge than completely shedding their energy into 
free quanta, for details and relevant references, see \cite{mazumdar02}.

Given an inflaton coupling to fermions of the type
$ h\phi\bar\psi\psi$ it has been shown that \cite{enqvist02a,enqvist02b}
reheating is driven by surface evaporation of inflatonic $Q$-balls for
relatively large Yukawa couplings $h\le 1$. In general $K$ and $h$ are
not independent quantities but are related to each other by
$|K| \sim C(h^2/16\pi^2)$. If the inflaton sector does not
belong to the hidden sector, it is very natural that the inflaton
coupling to other matter fields is relatively large, i.e.
$h\geq (m/M_{\rm P})$. In this regime the evaporation rate is
saturated and given by~\cite{enqvist02b}
\begin{equation}
    \label{evap2}
    \Gamma_Q = \frac{1}{Q}\frac{dQ}{dt}
        \simeq \frac{3}{16\pi \beta\zeta^2\gamma^2|K|^{3/2}}
                \left(\frac{m}{M_{\rm P}}\right)^2 m\,.
\end{equation}
Note that the decay rate is determined by the ratio
$m/M_{\rm P} \simeq 10^{-6}$, which is fixed by the anisotropies
seen in the cosmic microwave background radiation. Even
though we are in a relatively large coupling limit, the decay rate
mimics that of a Planck suppressed interaction of the inflatonic
$Q$-ball with matter fields. This is the most important feature of 
the decay rate of an inflatonic $Q$-ball evaporation.

Fermionic preheating~\cite{heitmann99} is not actually a problem 
in this case because the whole inflaton energy is never transferred 
during fermionic preheating because of Fermi-blocking, and the energy 
density stored in the fermions remains small compared to the inflaton 
energy density, as argued in~\cite{enqvist02b}. Fermions cannot scatter 
inflaton quanta off the condensate~\cite{heitmann99}, unlike in the case of
bosonic preheating~\cite{kofman94}. Note that inflatonic $Q$-ball production
is reminiscence to the bosonic preheating with a self coupling interaction 
appearing from the running mass of the inflaton. 

In conclusion, we have shown the formation of $Q$ and anti-$Q$-balls from
the fragmentation of the inflaton quanta. We have shown that surface 
evaporation of these $Q$-balls naturally gives rise to Planck suppressed 
decay rate inspite of the fact that the inflaton coupling to the fermion
is strong. We advocate here that our scenario can solve gravitino problem
via decreasing the reheating temperature for high scale inflation model,
see for details \cite{enqvist02a,enqvist02b}.

\vspace{-.9truecm}
\section*{Acknowledgments}

The author would like to thank Kari Enqvist and Shinta Kasuya for the
collaboration and for many useful discussion. 

\vspace{-.7truecm}


\begin{references}
\vspace{-1.5truecm}
\bibitem{liddle-lyth00}
A. R. Liddle, and D. H. Lyth, {\it Cosmological Inflation and
Large-Scale Structure}, Cambridge, Cambridge University Press (2000).

\bibitem{albrecht82}
A. Albrecht, P. J. Steinhardt, M. S. Turner, and F. Wilczek, Phys. Rev. Lett.
{\bf 48}, 1437 (1982); A. D. Dolgov, and A. D. Linde, Phys. Lett. B {\bf 116},
329 (1982); L. F. Abbott, E. Fahri, and M. Wise Phys. Lett. B {\bf 117}, 29
(1982).

\bibitem{kolbturner90}
E. W. Kolb, and M. S. Turner, {\it The Early Universe}, Adison-Wesley (1990).

\bibitem{traschen90}
J. Traschen, R. Brandenberger, Phys. Rev. D {\bf 42 }, 2491 (1990);
Y. Shtanov, Ukr. Fiz. Zh. {\bf 38}, 1425 (1993). (in Russian);
Y. Shtanov, J. Traschen, and R. Brandenberger, Phys. Rev. D {\bf 51},
5438 (1995).
%
\bibitem{kofman94}
L. Kofman, A. Linde, A. Starobinsky, Phys. Rev. Lett. {\bf 73}, 3195 (1994).
%
\bibitem{enqvist02a}
K. Enqvist, S. Kasuya, and A. Mazumdar, Phys. Rev. Lett. {\bf 89}, 091301
(2002).
%
\bibitem{enqvist02b}
K. Enqvist, S. Kasuya, and A. Mazumdar, Phys. Rev. D {\bf 68}, 043505
(2002).

\bibitem{maz}
A. Mazumdar, K. Enqvist, and S. Kasuya, hep-ph/0210241, {\it Talk given at
String Phenomenology, Oxford}. 

\bibitem{cohen86}
A. Cohen, S. Coleman, H. Georgi, and A. Manohar, Nucl. Phys. B {\bf 272}, 301
(1986).


\bibitem{mazumdar02}
K. Enqvist, and A. Mazumdar, {\it Cosmological Consequences Of MSSM Flat Directions},
hep-ph/0209244.

\bibitem{allahverdi02}
R. Allahverdi, and A. Mazumdar, {\it Nonthermal Leptogenesis With Almost Degenerate 
Superheavy Neutrinos}, To be published in Phys. Rev. D, hep-ph/0208268
 

\bibitem{ellis85b}
J. Ellis, D. V. Nanopoulos, and S. Sarkar, Phys. Lett. B {\bf 167}, 457 (1986).

\bibitem{maroto99}
A. L. Maroto, and A. Mazumdar, Phys. Rev. Lett. {\bf 84}, 1655 (2000);
R. Kallosh, L. Kofman, A. Linde, and A. Van Proyen,
Phys. Rev. D  {\bf 61}, 103503 (2000); M. Bastero-Gil, and A. Mazumdar, 
Phys. Rev. D {\bf 62}, 083510 (2000).

\bibitem{nilles01}
H. P. Nilles, K. A. Olive, and M. Peloso, Phys. Lett. B {\bf 522}, 304 (2001);
R. Allahverdi, K. Enqvist, and A. Mazumdar, Phys. Rev. D {\bf 65}, 103519
(2002).

\bibitem{allahverdi01}
R. Allahverdi, M. Bastero-Gil, and A. Mazumdar, Phys. Rev. D {\bf 64},
023516 (2001).

\bibitem{lythstewart}
D. H. Lyth, and E. D. Stewart, Phys. Rev. Lett. {\bf 75}, 201 (1995).

\bibitem{enqvist02c}
K. Enqvist, S. Kasuya, and A. Mazumdar, {\it Adiabatic Density Perturbations Aand Matter Generation From The MSSM}. hep-ph/0211147.

\bibitem{mazumdar99}
A. Mazumdar, Phys. Lett. B {\bf 469}, 55 (1999);
A. M. Green, and A. Mazumdar, Phys. Rev. D {\bf 65}, 105022 (2002). 

\bibitem{enqvist99}
K. Enqvist, and J. McDonald, Nucl. Phys. B {\bf 538}, 321 (1999).

\bibitem{kasuya0061}
S. Kasuya, and M. kawasaki, Phys. Rev. D {\bf 61}, 041301 (2000).

\bibitem{enqvist0163}
K. Enqvist, A. Jokinen, T. Multam\"aki, and I. Vilja, Phys. Rev. D {\bf 63},
083501 (2001).

\bibitem{heitmann99}
J. Baacke, K. Heitmann, C. Patzold, Phys. Rev. D {\bf 58}, 125013 (1998);
P.B. Greene, and L. Kofman, Phys. Lett. B {\bf 448}, (1999).


\end{references}
\end{document}